\documentclass{appolb}
\newcommand{\lsim}{\raisebox{-0.13cm}{~\shortstack{$<$ \\[-0.07cm] $\sim$}}~}
\newcommand{\gsim}{\raisebox{-0.13cm}{~\shortstack{$>$ \\[-0.07cm] $\sim$}}~}

\newcommand{\ra}{\rightarrow}

\newcommand{\ee}{e^+e^-}

\newcommand{\s}{\smallskip}
\newcommand{\nn}{\noindent}
\newcommand{\non}{\nonumber}
\newcommand{\beq}{\begin{eqnarray}}
\newcommand{\eeq}{\end{eqnarray}}

\usepackage{graphics,cite,amssymb,epsfig,float,psfrag,rotating}
\usepackage[usenames,dvips]{color}
\begin{document}
\title{Tests of the Higgs properties at the next colliders%
\thanks{Presented at  PLC2005 Warsaw and Kazimierz Lectures, 
5/09--08/09 2005.}%
}
\author{Abdelhak DJOUADI
\address{LPT, Universit\'e Paris--Sud, F--91405 Orsay, France}
}
\maketitle
\begin{abstract}
We discuss the tests of the fundamental properties of the Standard Model
Higgs boson that can be performed in the next round of experiments.
 \end{abstract}
  
\section{Introduction} 

The search for Higgs bosons is the primary mission of present and future
high--energy colliders. Detailed theoretical and experimental studies performed
in the last few years, have shown that the single neutral Higgs boson that is
predicted in the Standard Model (SM) \cite{Revue} could be discovered at the
upgraded Tevatron, if it is relatively light and if enough integrated luminosity
is collected \cite{Tevatron,Houches} and can be detected at the LHC
\cite{Houches,LHC} over its entire mass range 115 GeV $\lsim M_H \lsim 1$ TeV
in many redundant channels.

Should we then declare that we have done our homework and wait peacefully for 
the LHC to start operation? Well, discovering the Higgs boson is not the entire
story, and another goal, just as important, would be to probe the electroweak
symmetry breaking mechanism in all its facets. Once the Higgs boson is found,
the next step would therefore be to perform very high precision measurements to
explore all its fundamental properties. To achieve this goal in great detail,
one needs to measure all possible cross sections and decay branching ratios of
the Higgs bosons to derive their masses, their total decay widths, their
couplings to the other particles and their self--couplings, their spin--parity
quantum numbers, etc.  This needs very precise theoretical predictions and
more involved theoretical and experimental studies. In particular, all possible
production and decay channels of the Higgs particles, not only the dominant and
widely studied ones allowing for clear discovery, should be investigated.  This
also requires complementary detailed studies at future $\ee$ linear colliders,
where the clean environment and the expected high luminosity allow for very
high precision measurements \cite{e+e-,Tesla}. 

In this talk, I summarize the studies of the Higgs profile that can be 
performed at the LHC and ILC. Most of the material presented here relies on the
detailed discussion given in Refs.~\cite{Revue} to which we refer  for details.

\section{Higgs production and tests at the LHC} 

\nn The production mechanisms for the SM Higgs boson at hadron colliders are
\vspace*{-5mm}
\begin{eqnarray}
\begin{array}{lccl}
(a) & \ \ {\rm gluon~gluon~fusion} & \ \ gg  \ \ \ra & H \nonumber \\
(b) & \ \ {\rm association~with}~W/Z & \ \ q\bar{q} \ \ \ra & V + H \nonumber\\
(c) & \ \ WW/ZZ~{\rm fusion}       & \ \ VV \  \ra &  H \nonumber \\
(d) & \ \ {\rm association~with~}Q\bar{Q} & gg,q\bar{q}\ra & Q\bar{Q}+H
\nonumber
\end{array}
\end{eqnarray}
\vspace*{-4mm}

The cross sections are shown in Fig.~1 for the LHC with $\sqrt{s}=14$ TeV. Let
us briefly discuss the main features of each channel at the LHC [for the 
various Higgs decay  modes; see Ref.~\cite{Decays} for details]:

\begin{figure}[!h]
\begin{center}
\vspace*{-.6cm}
\psfig{figure=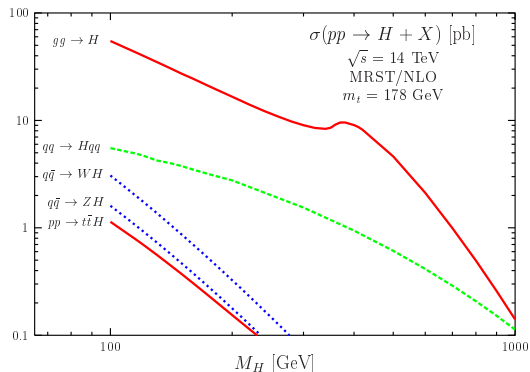,width=9cm}
\end{center}
\vspace*{-8.cm}
\caption[]{The Higgs boson production cross sections at the LHC in the 
dominant mechanisms as functions of $M_H$; from Ref.~\cite{Revue}.}
\vspace*{-2mm}
\end{figure}

{\bf a)} The dominant process, up to masses $M_H \lsim  700$ GeV, is by far the
$gg$ fusion  mechanism. The most promising signals are $H \to \gamma \gamma
(WW^*)$ in the mass range below 130 GeV (arround 160 GeV); for larger masses
it is $H \ra Z Z^{(*)} \ra 4 \ell^\pm$, with $\ell=e,\mu$, which from $M_H \gsim
500$ GeV can be complemented by $H \to ZZ \to \nu\bar{\nu} \ell^+ \ell^-$ and 
$H \to WW \to \nu \ell jj$. The higher order QCD corrections should  be taken
into account since they lead to an increase of the  cross sections by a factor
of $\sim 2$; see Ref.~\cite{Revue} for a review. 

{\bf b)} The $WW/ZZ$ fusion mechanism has the second largest cross section  [a
few picobarns for $M_H \lsim 250$ GeV] and rather small backgrounds [comparable
to the signal] allowing precision measurements, one can use forward--jet
tagging, mini--jet veto for low luminosity, and one can trigger on the central
Higgs decay products \cite{Dieter}. It has been shown that at least the decay $H
\to \tau^+ \tau^-$ as well as $H \to \gamma \gamma , WW^*$ can be detected and
could allow for Higgs coupling measurements \cite{Houches,Dieter}. 

{\bf c)} The associated production with gauge bosons, with $H \to b\bar{b}$  is
 the most relevant mechanism at the Tevatron \cite{Tevatron} but at the LHC,
 this process [also with the decays $H \to \gamma \gamma$] is challenging
 and needs a large luminosity.  

{\bf d)} Finally, Higgs boson production in association with top quarks, with $H
\to \gamma \gamma$ or $b\bar{b}$, could be observed at the LHC with  sufficient
luminosity and would allow the measurement of the important top Yukawa coupling.

Let us now turn to the measurements that can be performed at the LHC. We will
mostly rely on the summaries given in  Refs.~\cite{Revue,Houches,Dieter,LHC-LC}.
In most cases, a large, ${\cal O} (200)$ fb$^{-1}$, integrated
luminosity is assumed. 

$\bullet$ The Higgs boson mass can be measured with a very good accuracy. For
$M_H \lsim 400$ GeV, where $\Gamma_H$  is not too  large, a precision of
$\Delta M_H/M_H \sim 0.1$\% can be achieved in $H \to ZZ^{(*)} \to 4\ell^\pm$. 
In the ``low--mass" range, a slight improvement can be obtained by  considering
$H \to \gamma \gamma$. For $M_H \gsim 400$ GeV, the precision starts to
deteriorate because of the smaller rates. However, a precision of the order of
1\% can still be obtained up to $M_H\sim 800$ GeV if theoretical errors, such
as width effects, are not taken into account.  

$\bullet$ Using the same process, $H \to ZZ^{(*)} \to 4\ell^\pm$, the total
Higgs width can be measured for masses above $M_H \gsim 200$ GeV, when it is
large enough. While the precision is very poor near this mass value [a factor 
of 2], it improves to reach the level of $\sim 5$\% around $M_H \sim 400$ 
GeV. Here again, the theoretical errors are not taken into account.  

$\bullet$ The Higgs boson spin can be measured by looking at angular
correlations between the fermions in the final states in $H \to VV \to 4f$.  
However the cross sections are rather small and the environment
too difficult.  Only the measurement of the decay planes of the two $Z$ bosons
decaying into four leptons seems promising.  The Higgs CP properties and the
structure of the $HVV$ coupling can be also determined in the  fusion process,
$qq \to qqH$,  by looking at the azimuthal dependence of the two outgoing
forward tagging jets. The analysis is independent of the Higgs mass  and decay
modes but might be difficult because of background problems. 

$\bullet$ The direct measurement of the Higgs couplings to gauge bosons and
fermions is possible, but with rather poor accuracy. This is due to the limited
statistics, the large backgrounds, and the theoretical uncertainties from the
parton densities and the higher--order radiative corrections. To reduce some
uncertainties,  it is more interesting to measure ratios of cross sections where
the   normalization cancels out.  The cross sections times branching ratios
which can be measured in various channels at the LHC are shown in the  left-hand
side of  Fig.~2  for Higgs masses below 200 GeV \cite{Dieter}. A statistical 
precision of the order of 10 to 20\% can be achieved in some channels, while the
vector boson fusion process, $pp \to H qq \to W W qq$, leads to accuracies of
the order of a few percent.  Under some assumptions, these $\sigma \times  {\rm
BR}$ can be translated into Higgs partial widths in the various decay channels
$\Gamma_X \equiv \Gamma( H\to XX)$ \cite{Dieter}, which are proportional to the
square of the Higgs couplings, $g_{HXX}^2$. The expected accuracies, at the
level of 10 to 30\%,  are shown in the right--hand side of Fig.~2. One can
indirectly measure the total Higgs width $\Gamma$, and thus derive the absolute
values of the partial widths $\Gamma_X$, by making additional assumptions,
besides $g_{HWW}/g_{HZZ}$ universality

\begin{figure}[!h]
\vspace*{-2mm}
\begin{center}
\includegraphics[width=7.0cm,angle=90]{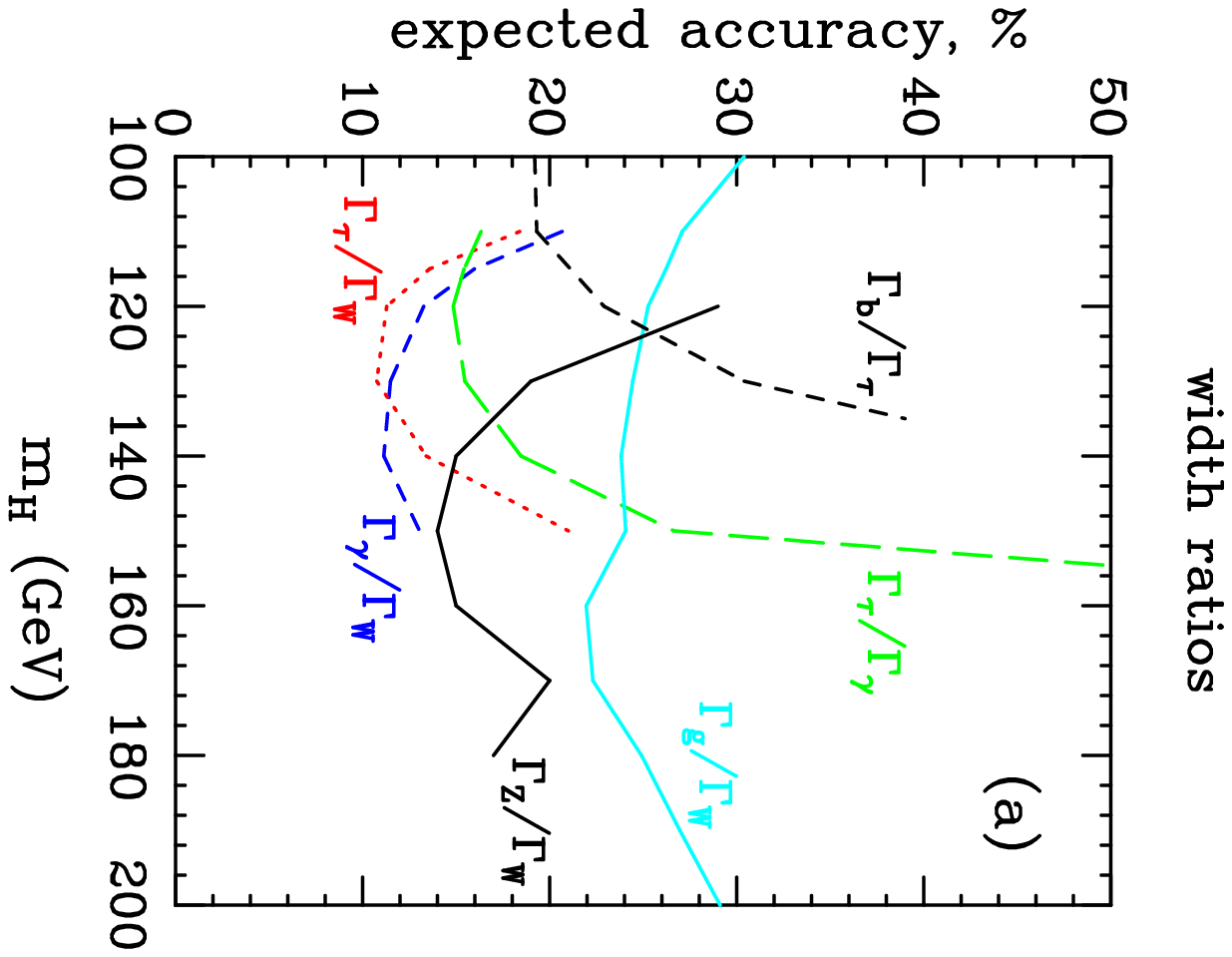} \hspace*{1cm}
\includegraphics[width=7.0cm,angle=90]{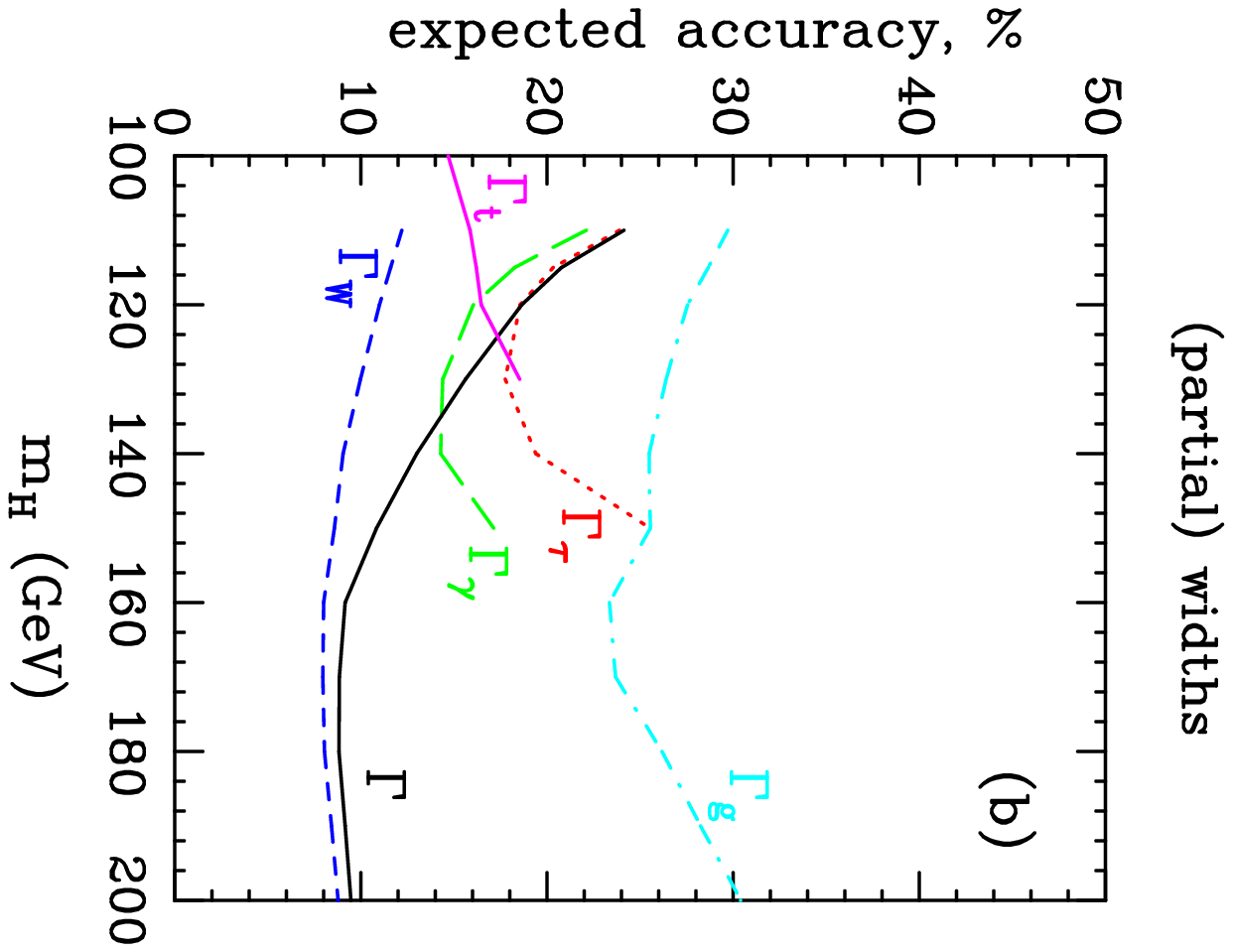}
\end{center}
\vspace*{-.5cm}
\caption[]{Relative accuracy expected at the LHC with a luminosity of 
200 fb$^{-1}$ for various ratios of Higgs partial widths (left) and 
indirect determination of partial and total widths $\tilde\Gamma_i$ and 
$\Gamma$ (right); from Ref.~\cite{Dieter}.} 
\vspace*{-.2cm}
\end{figure}

$\bullet$ The trilinear Higgs boson self--coupling $\lambda_{HHH}$ is too
difficult to be measured at the LHC because of the smallness of the $gg\to HH$
[and a fortiori the $VV \to HH$ and $qq \to HHV$] cross sections and the very
large backgrounds; see Ref.~\cite{Revue,LHC-LC} for a recent work and
references.

Note that some of the measurements discussed previously would greatly benefit
from an increase of the LHC luminosity (SLHC) or from an increase of the energy
(VLHC); see Ref.~\cite{SLHC+VLHC} for details.

\section{Higgs production and tests at the ILC} 

At $\ee$ linear colliders operating in the 300--1000 GeV energy range,  the main
production mechanisms for SM--like Higgs particles are 
\begin{eqnarray} 
\begin{array}{lccl} 
(a)  & \ \ {\rm Higgs\!-\!strahlung \ process} & \ \ \ee & \ra (Z) \ra Z+H \non \\ 
(b)  & \ \ WW \ {\rm fusion \ process} & \ \ \ee & \ra \bar{\nu} \ \nu \ (WW) 
\ra \bar{\nu} \ \nu \ + H \non \\ 
(c)  & \ \ ZZ \ {\rm fusion \ process} & \ \ \ee & \ra e^+ e^- (ZZ) \ra 
e^+ e^- + H \non \\ 
(d)  & \ \ {\rm radiation~off~tops} & \ \ \ee & \ra (\gamma,Z) \ra t\bar{t}+H 
\non \\ 
(e)  & \ \ {\rm double~Higgs\!-\!strahlung} & \ \ \ee & \ra (Z) \ra Z+HH 
\non 
\end{array} 
\end{eqnarray}

The Higgs--strahlung cross section scales as $1/s$ and therefore dominates at
low energies, while the $WW$ fusion mechanism  has a cross section that rises
like $\log(s/M_H^2)$ and dominates at high energies. At $\sqrt{s} \sim 500$ GeV,
the two processes have approximately the same cross sections, ${\cal O}
(100~{\rm fb})$ for the interesting range 100 GeV $\lsim M_H \lsim$ 200 GeV, as
shown in Fig.~3.With $ {\cal L} \sim 500$ fb$^{-1}$, as it was expected e.g.\ in
the TESLA design \cite{Tesla}, approximately 25.000 events per year can be
collected in each channel for a $M_H \sim 150$ GeV. This sample is more than
sufficient to discover the Higgs boson and to study its properties in detail.
Higgs masses of the order of 80\% of the c.m. energy can be probed, which means
that a 800 GeV collider can cover almost the entire mass range in the SM,  $M_H
\lsim 650$ GeV. The subleading processes (c)--(e) have orders of magnitude
smaller rates but they are very important for precision tests of  the Higgs
properties  as discussed below. 

\begin{figure}[h]
\vspace{-1.1cm}
\begin{center}
\hspace{-1.5cm}
\includegraphics[width=13.cm]{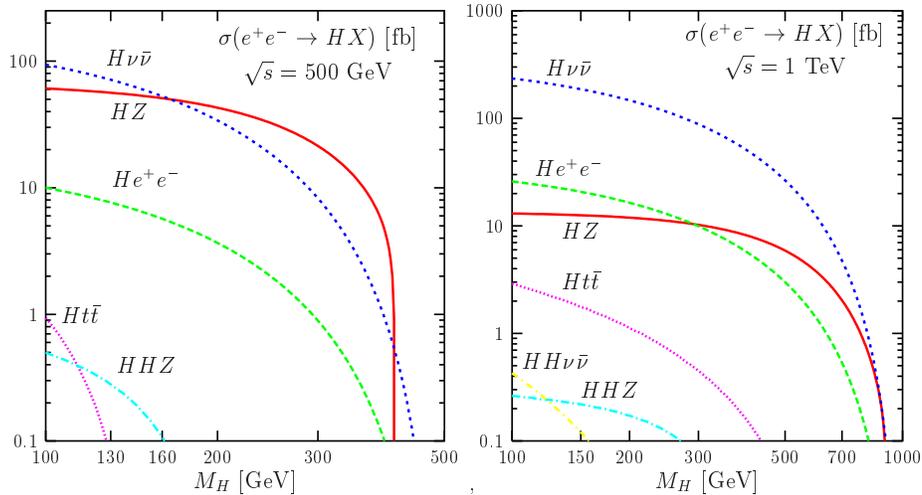}
\end{center}
\vspace{-11.4cm}
\caption[]{Production cross sections of the SM Higgs boson in $e^+ e^-$
collisions in the main and subdominant processes at $\sqrt{s}=500$ GeV (left) 
and 1 TeV (right) \cite{Revue}.}
\vspace{-.2cm}
\end{figure}

The determination of the Higgs properties can be done in great  detail in the
clean environment of the ILC \cite{e+e-,Tesla}.  In the following, relying on
analyses done for TESLA \cite{Tesla} [where the  references for the original
studies can be found], we   summarize the measurements in the case of the SM
Higgs.

$\bullet$ The measurement of the recoil $f\bar{f}$ mass in the strahlung
process, $\ee \ra ZH\ra H f\bar{f}$ allows a very good determination of $M_H$: 
at $\sqrt{s}=350$ GeV  with  ${\cal L}= 500$ fb$^{-1}$, a precision of $ \Delta
M_H \sim 50$ MeV can be reached for $M_H \sim 120$ GeV.  Accuracies  $\Delta M_H
\sim 80$ MeV can also be reached for $M_H=150$ and 180 GeV when the Higgs decays
mostly into gauge bosons.   

$\bullet$ The angular distribution of the $Z/H$ in the strahlung process, 
$\propto  \sin^2\theta$ at high energy, characterizes the production of a 
$J^P=0^+$ particle. The Higgs spin--parity quantum numbers can also be checked
by looking at correlations in the production $\ee \ra HZ \ra 4f$ or decay $H \ra
WW^* \ra 4f$ processes, as well as in the channel $H \ra \tau^+ \tau^-$ for $M_H
\lsim 140$ GeV. An unambiguous test of the Higgs CP nature  can be made  by
analyzing the spin--correlation in $H \to \tau^+ \tau^¯$ and possibly in  the
process $\ee \ra t \bar{t}H$ [or at laser photon colliders in the  loop--induced
process $\gamma \gamma \ra H$].  

$\bullet$ The Higgs couplings to $ZZ/WW$ bosons [which are predicted to be
proportional to the masses] can be directly determined by measuring the
production cross sections in the $\ee \ra H \ell^+ \ell^-$ and $H\nu \bar{\nu}$
processes. A precision less than $\sim$ 3\% at $\sqrt{s}\sim 500$ GeV with
${\cal L}= 500$ fb$^{-1}$ can be achieved. This leads to an accuracy of  $\lsim$
1.5\% on the $HVV$ couplings.  

$\bullet$ The measurement of the Higgs branching ratios (BRs) is of utmost
importance. For $M_H \lsim 130$ GeV a large variety of ratios can be measured:
the $b\bar{b}, c\bar{c}$  and $\tau^+ \tau^-$ BRs allow us to derive the
relative Higgs--fermion couplings and to check the prediction that they are
proportional to the masses. BR$(gg)$ is sensitive to the $t\bar{t}H$ coupling
and to new strongly interacting particles. BR$(WW)$ allows an indirect
measurement of the $HWW$ coupling, while BR($\gamma \gamma$) is also  important
since it is sensitive to new particles. Measurements at the level of a few
percent for most of the BRs at $M_H \sim 130$ GeV can be made [except for
$\gamma \gamma$ and $ZZ^*$ where the errors are of ${\cal O}(20\%)$]. 

$\bullet$ The Higgs coupling to top quarks, which is the largest coupling in the
SM, is directly accessible in the $\ee \ra t\bar{t}H$ process, although the
rates are low [see Fig.~3].  For $M_H \lsim 130$ GeV, $g_{Htt}$  can
be measured with a precision of less than 5\% at $\sqrt{s}\sim 800$ GeV with 
$ {\cal L} \sim 1$ ab$^{-1}$. 

$\bullet$ The total width of the Higgs boson, for masses less than $\sim 200$
GeV, is so small that it cannot be resolved experimentally. However, the
measurement of BR($H \ra WW$) allows an indirect determination of $\Gamma_H$,
since the $HWW$ coupling can be determined from the measurement of the Higgs
cross section in the $WW$ fusion process. [$\Gamma_{\rm tot}$ can
also be derived by measuring the $\gamma \gamma \to H$ cross section at a 
$\gamma\gamma$ collider or BR($H \to \gamma \gamma)$ in $\ee$].

$\bullet$ Finally, the measurement of the trilinear Higgs self--coupling, which
is the first non--trivial test of the Higgs potential, is accessible  in the
double Higgs production processes $\ee \ra ZHH$ [and in the $\ee \ra \nu
\bar{\nu}HH$ process at high energies]. Despite its smallness [see Fig.~3], the
cross  section can be determined with an accuracy of the order of 20\% at a 500
GeV  collider if a high luminosity, ${\cal L} \sim 1$ ab$^{-1}$, is
available. 

An illustration of the experimental accuracies that can be achieved in the
determination of the mass, CP--nature, total decay width and the various
couplings of the Higgs boson for $M_H=120$ and 140 GeV is shown in Tab.1 for
$\sqrt{s}=350$ GeV [for $M_H$ and the CP nature] and $500$ GeV [for
$\Gamma_{\rm tot}$ and all couplings except for $g_{Htt}$] and for $\int {\cal
L}=500$ fb$^{-1}$ [except for $g_{Htt}$ where $\sqrt{s}=1$ TeV and $\int {\cal
L}=1$ ab$^{-1}$ are assumed].

\begin{table}[htbp] 
\vspace*{-4mm} 
\begin{center}
\renewcommand{\arraystretch}{1.2}
\begin{tabular}{|c|c|c|c|}\hline 
$M_H$ (GeV) & $\Delta M_H$ & $\Delta {\rm CP}$ & $\Gamma_{\rm tot}$ \\ \hline 
$120$ & $\pm 0.033$ & $\pm 3.8$ & $\pm 6.1$\\ \hline 
$140$ & $\pm 0.05$ & $-$ & $\pm 4.5$\\ \hline 
\end{tabular}   
\vspace*{2mm}
\begin{tabular}{|c|c|c|c|c|c|c|c|c|}\hline 
$M_H$ (GeV) & $g_{HWW}$ & $g_{HZZ}$ & $g_{Htt}$ & $g_{Hbb}$ & $g_{Hcc}$ & 
$g_{H\tau \tau}$ & $g_{HHH}$  \\ \hline 
$120$ & $\pm 1.2$ & $\pm 1.2$ &  $\pm 3.0$ & $\pm 2.2$
& $\pm 3.7$ & $\pm 3.3$ & $\pm 17$  \\ \hline 
$140$ & $\pm 2.0$ & $\pm 1.3$ & $\pm 6.1$ & $\pm 2.2$ & $\pm 10$ & $\pm 4.8$ &
$\pm 23$  \\ \hline 
\end{tabular}
\vspace*{-2mm} 
\end{center} 
\vspace*{-2mm} 
\caption{Relative accuracies (in \%) on the Higgs boson mass, its CP mixture and
total width (top) and on its couplings (bottom) obtained at TESLA with 
$\sqrt{s}=350,500$ GeV and $\int {\cal L}=500$ fb$^{-1}$  (except for top); 
Ref.~\cite{Tesla}. }
\vspace*{-2mm}
\end{table}

\section{Conclusions}

The detection of a Higgs particle is possible at the upgraded Tevatron for $M_H
\lsim 130$ GeV and is not a problem at the LHC where even much heavier Higgs
bosons can be probed. Relatively light Higgs bosons can also be found at future
$\ee$ colliders with c.m.\,energies $\sqrt{s} \gsim 350$ GeV; the signals are
very clear, and the expected high luminosity allows a thorough investigation of
their fundamental properties.   Some of this discussion can  of course be
extended to the the lightest Higgs particle of the minimal SUSY extension of the
SM (MSSM), which is expected to have a mass smaller than 140 GeV; see 
Ref.~\cite{Mh}.   In fact, a very important issue once Higgs particles are
found, will be to probe in all its facets the electroweak symmetry breaking
mechanism.  In many aspects, the searches and tests at future $\ee$ colliders
are complementary to those that will be performed at the LHC.  An example can be
given in the context of the MSSM.

\begin{figure}[ht!]
\vspace*{-.2cm}
\begin{center}
\epsfig{file=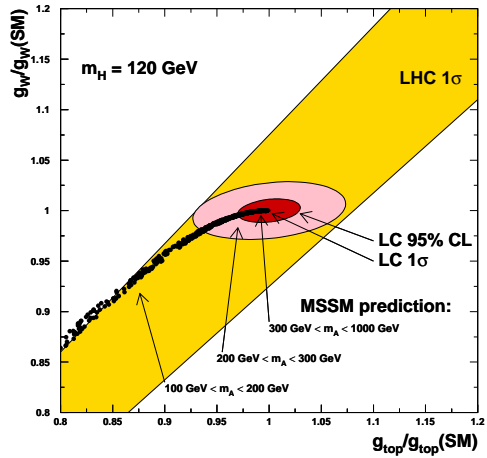,width=0.5\linewidth}
\vspace*{-.3cm}
\caption{Higgs boson coupling determinations at TESLA for $M_H=120$ GeV with 
500~fb$^{-1}$ of data, the $1\sigma$ LHC constraint and the expected 
deviations in the MSSM for various $M_A$ ranges are also shown; from Ref.~\cite{Tesla}. }
\end{center}
\label{fig:hfitter}
\vspace*{-.8cm}
\end{figure}

In constrained scenarios, such as the minimal supergravity model, the heavier
$H,A$ and $H^\pm$ bosons tend to have masses of the order of several hundred GeV
and therefore will escape detection at both the LHC and linear collider. In this
parameter range, the $h$ boson couplings to fermions and gauge bosons will be
almost SM--like and, because of the relatively poor accuracy of the measurements
at the LHC, it would be difficult to resolve between the SM and MSSM (or
extended) scenarios. At the ILC, the Higgs couplings can be measured with a
great accuracy, allowing a distinction between the SM and the MSSM Higgs boson
to be made close to the decoupling limit, i.e. for pseudoscalar boson masses,
which are not accessible at the LHC.  This is exemplified in  Fig.~4, where the
accuracy in the determination of the Higgs couplings to $t\bar{t}$ and $WW$ 
states are displayed at LHC and ILC, together with the predicted values in the
MSSM for different values of $M_A$.  At the ILC, the two scenarios can be
distinguished for pseudoscalar Higgs masses up to 1 TeV and, thus, beyond the
LHC reach.\s

{\bf Acknowledgements:} I thank the organizers of the conference, in particular
Maria Krawczyk, for the invitation and for the nice atmosphere. 

\vspace*{-2mm}


\begin{thebibliography}{199} 

\bibitem{Revue} For recent reviews on the Higgs sector in the SM and MSSM and a
complete set of references, see: A. Djouadi, hep-ph/0503172 and hep-ph/0503173. 

\bibitem{Tevatron} M. Carena et al., hep-ph/0010338.
 
\bibitem{Houches} Procs. of  Les Houches,  A. Djouadi et al., hep-ph/0002258; 
D. Cavalli et al., hep-ph/0203056; K. Assamagan et al., hep-ph/0406152.

\bibitem{LHC} CMS TR, CERN/LHCC/94-38; ATLAS TDR, CERN/LHCC/99-15;
G. Branson et al. (CMS and ATLAS),  Eur. Phys. J. direct C4 (2002) N1;
V. B\"usher and K. Jakobs, hep-ph/0504099. 

\bibitem{e+e-} E. Accomando, Phys. Rept. 299 (1998) 1;    American LC WG,  
hep-ex/0106057;  ACFA LC Higgs WG, hep-ph/0301172. 

\bibitem{Tesla} TESLA Technical Design Report, Part III, hep-ph/0106315. For
updates see,  K. Desch  et al., hep-ph/0311092, T. Barklow, hep-ph/0312268; 
J.-C. Brient, Note LC-PHSM-2002-03; S. Kanemura, hep-ph/0410133.   

\bibitem{Decays} A. Djouadi et al., Z. Phys. C70 (1996) 427; Z. Phys. C70 (1996)
435; Comput. Phys. Commun. 108 (1998) 56.

\bibitem{Dieter} D. Zeppenfeld et al., Phys. Rev. D62 (2000) 013009;  M.
D\"uhrssen et al. Phys. Rev. D70 (2004) 113009;  M. D\"uhrssen,  ATLAS Note
PHYS-2003-030.

\bibitem{LHC-LC} G. Weiglein et al. (LHC--LC Study Group), 
hep-ph/0410364.


\bibitem{SLHC+VLHC} F. Gianotti et al., hep-ph/0204087;  U. Baur et al.,
hep-ph/0201227.

\bibitem{Mh} For recent discussions, see 
G.~Degrassi et al., Eur. Phys. J. C28 (2003) 133; 
B. Allanach et al., JHEP 0409 (2004) 044.   


\end{thebibliography}
\end{document}